# Half a Century of Kinetic Solar Wind Models

Joseph Lemaire

*Institut d'Aéronomie Spatiale de Belgique (3, Avenue Circulaire, B-1180 Bruxelles, Belgium)*
*jfl@astr.ucl.ac.be*

**Abstract.** I outline the development of four generations of kinetic models, starting with Chamberlain[1]'s solar breeze exospheric model. It is shown why this first kinetic model did not give apposite supersonic evaporation velocities, like early hydrodynamic models of the solar wind. When a self-consistent polarization electric potential distribution is used in the coronal plasma, instead of the Pannekoek-Rosseland's one, supersonic bulk velocities are readily obtained in the second generation of kinetic models. It is outlined how the third and fourth generations of these models have improved the agreement with observations of slow and fast speed solar wind streams.

## INTRODUCTION

In 1961, Chamberlain[1] questioned the validity of Parker's[2] isothermal hydrodynamic model for the coronal expansion. He proposed an alternative description to model the expansion of the solar corona. His model is remembered as the 'solar breeze model'. But since it did not take into account the right polarization electric field distribution, it was unable to obtain the supersonic bulk speed observed at 1AU.

The development of coronal exospheric models using self-consistent electric field introduced independently by Jockers[7] and Lemaire and Scherer[4-6] will be reviewed. The second generation of kinetic models was based on Maxwellian velocity distribution functions (VDF) at the exobase, and monotonic radial distributions for the potential energy of protons and electrons.

Non-Maxwellian velocity distribution functions[8-10] and non-monotonic functions for the total potential energy[11-13] were introduced in 1996, to increase the potentiality of exospheric models. These form the third generation of kinetic models. They till belong to the category of "zero-order kinetic models", since collisions were ignored for all of them. In the fourth generation of kinetic models[14-17], Coulomb collisions have now been taken into account by solving the stationary Fokker-Planck equation for the solar wind (SW) electrons.

The development of these kinetic models enabled us to understand more clearly the physical mechanism by which the coronal protons are accelerated to supersonic speed; therefore they stand as unique alternatives of the many hydrodynamic models heartily developed since 1958.

An extensive review of kinetic and hydrodynamic SW models, and of their chronologic development will be found in Echim et al.[18]. The monograph by Meyer-Vernet[20] gives a seminal and comprehensive overview of the solar wind in general. See Marsch[19] for a review of wave-particle interactions in the SW.

## HYDRODYNAMIC MODELS

Fig. 1 illustrates the different families of SW hydrodynamic models. For convenience it was assumed that the coronal temperature is uniform and equal to $1.0 \times 10^6$ K. The left panel shows $u_r(h)$, the radial bulk velocity as a function of altitude. Two subsonic solutions are shown; the critical subsonic-supersonic solution passing through a singular point (the solid square at $h = 4\ R_S$), and two physically irrelevant stationary solutions of the Euler hydrodynamic equations used to calculate these solutions.

All subsonic solutions lead to excessive pressures and densities at infinity. Parker[2] pointed out that only the critical solution with supersonic speed at large distances leads to small enough kinetic pressures and densities at infinity, and is thus compatible with the conditions prevailing in the interstellar medium.

In 1961, Chamberlain[1] claimed that the single fluid hydrodynamic equations used by Parker[2] to model the solar wind plasma flow fail to be applicable beyond a heliocentric distance of about 2.5 $R_S$. He argued that beyond this radial distance the coronal plasma density becomes so small that its Knudsen number, *Kn*, becomes larger than unity. Indeed, when the Knudsen number of a gas exceeds unity the Euler approximation of the hierarchy of moment equations becomes questionable: there is then no valid justification to assume that the kinetic pressure tensor stays isotropic, and the heat flux equal to zero (in the adiabatic models) or infinite (in the isothermal ones).

The surface where *Kn = 1*, is called the exobase. It is the place where the mean free path (mfp) of particles becomes equal to the density scale height, $H = k(T_e+T_p)/2m_p\, g$. The right panel of Fig. 1 shows the distributions of *Kn* as a function of *h,* for the isothermal hydrodynamic solutions of the left panel. The dots on the dotted lines in both panels mark the altitudes where the hydrodynamic models become collisionless. Note that the exobase altitudes are in general located below the critical point.

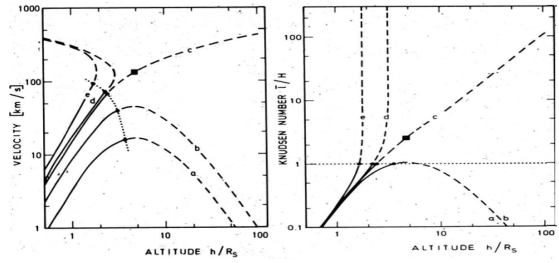

**FIGURE 1.** Isothermal hydrodynamic solar wind models. Left panel: expansion velocities versus altitude for the different classes of hydrodynamic models. Right panel: distributions of Knudsen number versus altitude in these hydrodynamic models (*adapted from Brasseur and Lemaire[21]*).

Since Parker's pioneering work, many more sophisticated one-fluid, two-fluid or multi-fluid SW models have been published. A description of this corpus of hydrodynamic SW models can be found in Aschwanden[22], and in the review by Echim et al.[18]. Only kinetic SW models will be briefly reviewed below. I will show how the successive generations of kinetic models have improved our understanding of the physical mechanism by which coronal protons are accelerated to supersonic speeds, and how they do provide results which are in rather fair agreement with SW observations at 1AU.

## THE SOLAR BREEZE MODEL

Adopting Jeans exospheric theory for planetary atmospheres Chamberlain[1] developed the first exospheric theory describing the evaporation of protons from the solar corona. In ion-exospheres the guiding centers of protons are free to move along magnetic field lines (assumed to be radial for convenience); some of the protons having large enough velocities are able to escape out of the gravitational well, presumably without being impeded neither by Coulomb collisions, nor by wave-particle interactions.

The protons evaporating from the corona are decelerated by the gravitational force, but they are accelerated by the ambient polarization electric field directed away from the Sun. This electric field is well known in ionospheric physics as the Pannekoek[23]-Rosseland[24] (PR) electric field. The ratio of the electric force and gravitational force acting on the major ion species (i.e. protons) is then determined as $|eE/(m_p - m_e)g| = 0.5$. This simple relationship between *E* and *g,* holds only when the electron and proton temperatures are equal, and when the plasma is in hydrostatic equilibrium. It can be verified that the total forces acting on protons and electrons are then precisely equal; the density scale height of electrons and protons are then equal, as required to keep the plasma quasi-neutral at all altitudes in the gravitational field *g*. The PR electric field is thus essential to prevent the heavy ions to diffuse with respect to electrons, despite their larger gravitational force.

It is this well known electric field that was implemented in the solar breeze model to keep the plasma quasi-neutral. Assuming the proton VDF to be a truncated Maxwellian at the exobase, with ballistic, trapped/captive and escaping protons, but none coming in from infinity, Chamberlain[1] developed analytical expressions for the escape flux of protons and for their densities in the exosphere. This enabled him to calculate an average velocity of evaporating protons. He found a value of 20 km/s at 1AU, for an exobase temperature of $2\times10^6$ K. This was much smaller than the supersonic speed predicted by the critical solution of the hydrodynamic SW model; it was much too small also compared to supersonic proton velocities consistently observed at 1AU since 1961.

The inability of the controversial solar breeze model to predict the observed supersonic SW velocity led exospheric models and kinetic approaches in disrepute. It was not before 1969 that Lemaire and Scherer [4-5] and Jockers[7] independently discovered the reason why Chamberlain[1] failed to obtain the apposite answer. The only reason was the unjustified implementation of the PR electric field in the solar breeze model. By 1969 it was recalled that the PR field holds exclusively when the ionized gas is in hydrostatic and diffusive equilibrium, but not when it is expanding as the SW or the solar breeze.

What was not noticed until then is that the critical escape energy of protons is equal to the critical escape energy of electrons, when the PR field is assumed. As a consequence, Jeans' escape flux of electrons evaporating from the exobase is 42 times larger than the escape flux of protons. Indeed, the Jeans' flux is proportional to the thermal velocity of the particles, and is therefore $(m_p/m_e)^{1/2}$ times larger for electrons than for protons if their temperatures are equal.

As a result of the larger escape flux of electrons, the corona will charge up until the electrostatic potential $\Delta\Phi_E$, has increased up to a value for which the escape flux of electrons is reduced to become precisely equal to that of protons. When the coronal electron and proton temperatures are $1.5 \times 10^6$ K the "zero-current condition" is satisfied when $\Delta\Phi_E = 600$ Volts. In the solar breeze $\Delta\Phi_E$, was only 150 Volts, since the PR field was implicitly adopted.

As a consequence of the enhanced value of $e\Delta\Phi_E$, the electrostatic potential energy of protons has become larger than their gravitational potential energy, $m_p\Delta\Phi_g$. This is why all protons can now escape out of the Sun's gravitational well, and gain supersonic velocities at large radial distances.

It is this larger polarization electric field that is implemented in exospheric models of the second generation which are outlined in the next section. With this change implemented in exospheric models, they are not less adequate and no less valuable than hydrodynamic SW models. Both approaches are complementary as two representations of the same reality.

## THE SECOND GENERATION OF EXOSPHERIC MODELS

Using an equatorial electron density distribution deduced by Pottasch[25] from eclipse observations, Lemaire and Scherer[6] determined the coronal temperatures of electrons and protons for which the exobase would be at the same altitude ($h_o$) for both species. In their models $h_o$ is a free input parameter which might be changed to obtain a range of values for the bulk velocities at 1AU. In order to obtain 320 km/s for the SW velocity of at 1AU (i.e. the average value observed in the slow SW[26,27]), the exobase altitude had to be fixed at $h_o = 5.6\ R_S$, where $n_o=3.1\ 10^4\ cm^{-3}$, $T_e=1.52\ 10^6 K$ and $T_p=0.984\ 10^6 K$.

Adopting truncated Maxwellians characterized by these densities and temperatures for the protons and electrons at the exobase Lemaire and Scherer[6] obtained the radial distributions illustrated in Fig. 2[*].

---

[*] By changing $h_o$ as well as $n_o$, $T_e$ and $T_p$, as explained by Lemaire and Scherer[6], a whole family of exospheric models can be generated. The values of the proton bulk velocities ($u_r$) and temperatures ($<T_p>$) thus generated are correlated according to an almost-quadratic relationship illustrated in Fig. 12 of Lemaire and Scherer[30]. This correlation happens to be similar to that observed in the quiet solar wind at 1 AU (see also Fig. 4 in Echim et al.[18]).

The solid line in panel "a" gives the electron and proton number densities above the exobase. Pottasch's[25] empirical coronal densities are displayed by squares for $r < 20\ R_S$, while the SW observations at 1 AU are given by the error bar at $215\ R_S$.

Panel "b" shows the distribution of the expansion velocity; it can be checked that at 1 AU $u_r=320$ km/s, corresponding to quiet or slow SW observations[26-27]. The electrostatic potential difference $\Delta\Phi_E$ is 670 Volts; as emphasized above it is this electric potential that accelerates the protons to this supersonic bulk velocity.

Panel "c" shows the radial distributions of electron and proton perpendicular temperatures; these parameters characterize their velocity dispersion in the direction perpendicular to the magnetic field lines (supposed to be radial). While these transverse temperatures are much smaller than those generally observed, however, the average temperatures $<T_p>$ and $<T_e>$ are in good agreement with those observed in the quiet or slow SW[26-29], as indicated by the error bars in panel "d".

The predicted ratios of the parallel and perpendicular temperatures or pressures are much larger than the observed ones. In their discussion section Lemaire and Scherer[6] attribute these excessive temperature anisotropies to the absence of collisions and wave-particle interactions in their exospheric models.

Part of the pitch angle anisotropy can be reduced by replacing the postulated radial interplanetary magnetic field line distribution by a spiral shaped ones. This was shown by Issautier et al.[10] and Pierrard et al.[34]. Nevertheless, this geometrical correction is unable to reduce the anisotropies enough to fit the observations at 1 AU.

Another restriction limits exospheric SW models: their inability to achieve bulk velocities exceeding 600-700 km/s, as measured at 1 AU in fast speed streams which originate from coronal holes. According to the second generation of exospheric models the SW originating from these colder regions should have a smaller bulk velocity at 1 AU, instead of a larger one. Note that a similar shortcoming plagues hydrodynamic models as well, unless in-situ heating is arbitrarily added to boost the SW to any higher bulk speed value.

This shortcoming prompted the teams of modelers at LESIA, Meudon, and BISA, Brussels, to search for new ways improving exospheric models, and increasing the SW expansion velocity without increasing the exobase temperature beyond acceptable limits. Two main adjustments were proposed in the 1990's. They led to the development of the third generation of SW exospheric models.

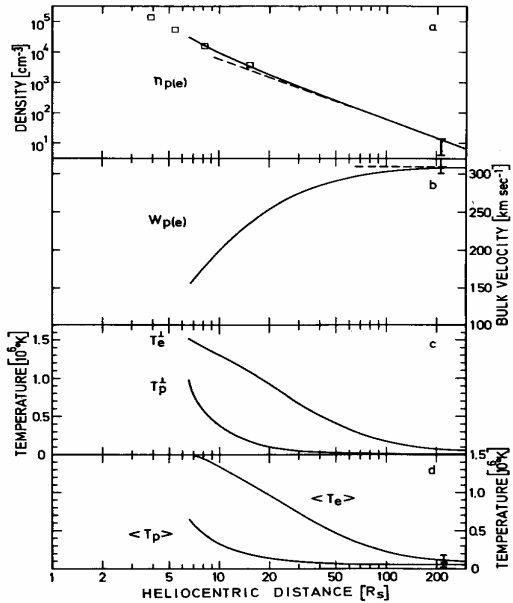

**FIGURE 2.** Lemaire-Scherer's exospheric model of the solar wind with an exobase at $r = 6.6\ R_S$ : $n_o = 3.1\ 10^4\ cm^{-3}$, $T_e = 1.52\ 10^6 K$ and $T_p = 0.984\ 10^6 K$ (adapted from Lemaire and Scherer[6,30]).

# THE THIRD GENERATION OF EXOSPHERIC MODELS

In the next generation of kinetic models, the VDF at the exobase were first assumed to be "kappa" or Lorentzian functions with a power-law distribution for supra-thermal particles. "Kappa" VDFs exhibit power law distributions, instead of exponential tails as Maxwellians functions; many energy spectra of charged particles in space have this particularity.

*Enhancing the Population of Supra-thermal Electrons*

The panels of Fig. 3 show the distributions of $u_r(r)$, as well as the sum of the gravitational and electrostatic potential energies of protons for three different types of exospheric models: (i) the thick curves correspond to a model of second generation with Maxwellian VDFs, and an exobase at $h_o = 6\ R_S$; it is similar to Lemaire-Scherer's model displayed in Fig. 2; (ii) the thin solid curves correspond to a third generation exospheric model with the same exobase altitude and temperature as in the previous one, but the VDFs are "kappa functions" with $\kappa = 3$. For this third generation model the value of $\Delta\Phi_E$ is larger, and consequently the bulk speed is larger (> 450 km/s).

This exercise shows the key role played by the supra-thermal electrons in accelerating the solar wind protons to supersonic velocity. Indeed, these electrons are, so to say with Parker[35] in this volume, "the 'horses' that drag the 'cart' (loaded with massive and positive ions), so that the two run away together" : i.e. at the same rate. It should be added that the role of the lower energy, ballistic and trapped electrons is to balance the charge density of the ions carried in the 'cart': in other words they mainly keep the exospheric plasma quasi-neutral.

*Lowering the Exobase Altitude*

Since plasma densities and temperatures are reduced in coronal holes, the mean free path of particles is necessarily larger than elsewhere. As a consequence, the exobase is likely to be located at a lower altitude in coronal holes. This implies that in the lower part of the exosphere the downward gravitational force, $m_p g$, is larger than the electric force, $eE$, which accelerates the protons upwards. In other words, at the base of the ion-exosphere $R_p(r)$, the total potential energy of protons, is dominated by the gravitational field, and it is still an increasing function of $r$, reaching a maximum value at $r = r_{max}$ where $|eE/m_p g|=1$. Beyond this heliocentric distance, $R_p(r)$ decreases with altitude, as in the second generation models for which the exobase was located higher up in the equatorial corona.

The exobase is at $r_0 = 1.1\ R_S$, and the maximum of $R_p(r)$ is located at $r_{max} = 1.9\ R_S$ for the model illustrated in Fig.3. Ballistic and trapped protons are present below $r_{max}$, since some of these ions don't have enough kinetic energy to overpass the potential barrier, $R_{p,max}$. Note that none of these trapped and ballistic protons contribute to the net SW flux which is thus significantly reduced. A similar reduction is required for the Jeans escape flux of electrons. As a consequence, a larger value for $\Delta\Phi_E$, is imposed by the zero-current condition.

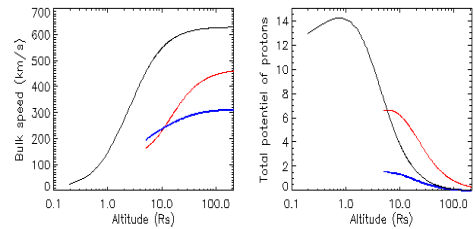

**FIGURE 3.** Comparison of exospheric solar wind models belonging to the second and third generations. Left panel: expansion velocity, $u_r(r)$. Right panel: total potential of protons, $R_p(r)$, normalized by $kT_{po}$ (adapted from Lamy et al.[12]).

To evaluate densities, fluxes, expansion velocity, pressure tensor components, temperatures, energy and heat fluxes in such more elaborated exospheric models, new analytical formulae have been developed by Lamy et al.[12] to calculate the moments of the VDFs above the exobase. An iterative method has to be used first to determine the numerical values of $r_{max}$ and $R_{p,max}$, the additional parameters of these new kinetic models. These third generation exospheric SW models have been implemented on the ESA public Website http://www.spaceweather.eu/he/kinetic_sw.

From the left panel of Fig. 3 it can be seen that for this new brand of kinetic models the SW velocity becomes larger than 600 km/s. Nevertheless, values as large as 900 km/s have not yet been achieved with this ultimate improvement of exospheric models for the solar wind expansion.

*Two Maxwellian Velocity Distribution Functions*

Another interesting study was published by Zouganelis et al.[13] where non-thermal VDFs have been used at the exobase, instead of truncated Maxwellians. This additional study confirms the key role played by the population of suprathermal electrons in the process of accelerating solar wind ions to higher terminal speeds. They demonstrate that higher terminal speeds can also be obtained by using a sum of two Maxwellian VDFs: i.e. the first Maxwellian electron population with normal coronal temperature representing the core electrons, and the second population of electrons of higher temperature, corresponding to the halo electrons.

Zouganelis et al.[13] confirmed also that, when the exobase is lower than $r_{max}$, the SW velocity at 1 AU increases when $h_o$ is decreasing, all other model parameters being unchanged. Furthermore, they showed that, in the case of non-monotonic proton potential profiles, the terminal SW velocity is anti-correlated with $T_{po}/T_{eo}$, the ratio of the proton to electron temperatures at the exobase[13].

Zouganelis et al.[33] compared terminal SW speeds obtained from their third generation exospheric models, with those obtained from numerical simulations taking into account Coulomb collisions. The bulk velocities at 1 AU obtained with and without Coulomb collisions are rather similar. This unexpected agreement might be attributed to the presence of trapped/captive electrons which were always postulated to be present in second and third generation exospheric models. Indeed, the Coulomb collision time required to scatter thermal electrons into such trapped trajectories is much smaller than the collision time of the protons, and also the time for the coronal plasma to expand up to 1AU.

# POTENTIALITIES OF EXOSPHERIC MODELS

Despite the rather artificial truncation of the VDFs, exospheric models offer the unprecedented advantage of giving clues to understand the physics of the solar wind acceleration without having to integrate a set of coupled non-linear differential equations across any mathematical singularity.

In single-fluid hydrodynamic formulations, the electrostatic force accelerating the protons to supersonic speeds is not evidenced: it is hidden in the single-fluid momentum equation within the gradient of the total kinetic pressure. This is probably why there are still complains that the physical mechanism accelerating the solar wind is not well understood.

In addition to the shortcoming of the solar breeze model for predicting the observed supersonic bulk speeds, there is another reason why exospheric models did not become popular. Indeed, it was erroneously considered that the moments of exospheric VDFs would not satisfy the hierarchy of moments equations which are derived from Boltzmann's equation[†]. But such a belief was a misbelief. Indeed, analytical expressions for the density, the bulk flow speed, the energy flux, the kinetic pressure tensor, and for all higher order moments of exospheric VDFs are exact solutions of the hierarchy of moments equations.

# HIGHER ORDER KINETIC MODELS

The excessive temperature anisotropies plaguing all exospheric models are partly attributable to the absence of Coulomb collision and wave-particle interactions above the exobase. This was argued by Lemaire and Scherer[6, 30], as well as by Griffel and Davis[31] who developed a BGK kinetic model with an ad hoc constant collision frequency intending to reduce these too large temperature asymmetries.

To eliminate this inherent shortcoming of collisionless SW models, higher-order kinetic models have been developed taking into account of the pitch angle scattering by Coulomb collisions in the exosphere. First attempts along this perspective have been worked out by Lie-Svendsen et al.[15] and by Pierrard et al[14, 16-17]. In these fourth generation kinetic models, stationary solutions of the Fokker-Planck equation have been calculated numerically. These authors obtained the velocity distribution function of SW electrons by two different mathematical methods: respectively, the finite difference method, and a spectral method based on a polynomial expansion of the VDF. This special expansion of the VDF in terms of 'speed polynomials' was adapted from Pierrard's Ph.D. thesis[14] on kinetic models for the polar wind.

---

[†] The moments equations correspond to exact transport equations, from which various approximations of the hydrodynamic equations have been derived by adopting different convenient closure methods (Chapman-Enskog's, Grad's….methods).

The solutions of the Fokker-Planck equation indicate that the pitch angle anisotropies of the halo and strahl electrons observed at 1 AU map down into the solar corona as a slight asymmetry of the coronal electron VDF. This is recalled in the paper by Pierrard and Voitenko[32]. In other words, the pitch angle asymmetry observed in the electron VDF at 1 AU is not generated by collisions or instabilities within the interplanetary plasma 'en route' to 1 AU.

## CONCLUSIONS AND PERSPECTIVES

The fourth generation of kinetic SW models should be further developed in future, by solving simultaneously coupled Fokker-Planck equations for the electrons, protons and possibly for other minor SW ions. Solving time dependent formulations of the Fokker-Planck equation, with and without in situ ionization/recombination or/and in situ heating processes, should be able to model also the propagation of shock waves or ICMEs (Interplanetary Coronal Mass Ejections) in the SW. These future efforts could complement the current MHD simulations, which are appropriate approximations for most space weather applications, but certainly not for an in-depth theoretical description of SW kinetic processes.

## ACKNOWLEDGMENTS

I wish to acknowledge fruitful collaboration with M. Scherer, V. Pierrard, H. Lamy and M. Echim from BISA, as well as a rewarding cooperation with N. Meyer-Vernet, K. Issautier, M. Maksimovic, and I. Zouganelis, from LESIA. I also appreciated discussing this material with E.N. Parker, Univ. Chicago, Bernie Shizgal, UBC, O. Lie-Svendsen, NDRSE, A. Zhukov's, ROB, and Y. Voitenko's, BISA. I wish to thank the referee for his constructive remarks.